# A half-wave voltage regulation method of electro-optic crystal based on center-symmetric electrodes and arbitrary direction electric field

**Nan Xie[1], Yifan Lin[1], Junfeng Chen[2], Qifeng Xu[1] and Yifan Huang[1]**
[1]College of Electric Engineering and Automation, Fuzhou University, Fuzhou, Fujian 350116, People's Republic of China
[2]Shanghai Institute of Ceramics, Chinese Academy of Sciences, Shanghai 201899, People's Republic of China

Corresponding author: Qifeng Xu (e-mail: ranger123098@163.com).

This work was supported by the National Natural Science Foundation of China under Grant Nos. 51977038 and Nos. 51807030

**ABSTRACT** To solve the problems of instability and thermal stress birefringence of voltage divided medium in optical voltage sensor (OVS), a half-wave voltage regulation method based on a central symmetry electrode and an arbitrary directional electric field is proposed in this paper. The method uses a center-symmetric copper foil electrode or the ITO (Indium Tin Oxide) transparent electrode to apply voltage, which can arbitrarily adjust the direction of the average electric field in the crystal, so as to improve the half-wave voltage of the bismuth germanate ($Bi_4Ge_3O_{12}$, or BGO) crystal directly without additional $SF_6$ gas or quartz glass. The internal electric field of the BGO crystal is simulated by the finite element method, and the half-wave voltage is calculated by the coupling wave theory of electro-optic effect. The results show that the maximum regulated half-wave voltage, considering the 1° angle, can reach 2.67e3 kV. And this is consistent to the experimental results 1.72e3 kV with difference of 35.6%. For the measurement of 110kV line voltage, that is, 63.5kV phase voltage, to ensure a 0.2 accuracy class, the angle error of incident light, and the polarization direction and the electric field direction is required to be less than 3.9', which can be achieved by the existing technologies.

## I. Introduction

The BGO crystal [1] is commonly used in optical voltage sensor [2-4], which belongs to the cubic crystal system $\bar{4}3m$ point group, and has no natural linear birefringence and circular birefringence. It is transparent in visible and near-infrared wavelengths region, and is available in a large size with a good optical quality. And it has no piezoelectric effect and small temperature coefficient.

For 110kV optical voltage sensor, a quartz glass voltage divider or $SF_6$ gas voltage divider is usually used to increase the half-wave voltage (HWV) of the whole system, and realize the approximate linear measurement of power frequency voltage within a small angle phase delay. Hofmann Christensen et al. [5] reported an OVS based on transverse modulation, with $SF_6$ gas voltage divider. BGO crystal are pasted with two right-angle prisms and fixed in the center of the ground electrode. There is a large gap between the high voltage electrode and BGO crystal, which is filled with $SF_6$ gas. Santos et al. reported an OVS based on longitudinal modulation, using multiple quartz media as voltage divider [6, 7]. Based on this work, our research group proposed a new OVS based on quartz glass divider medium to improve the electric field uniformity along the optical path [8]. In addition, we proposed the longitudinal modulation OVS based on the medium wrapping method to improve the electric field distributions in the BGO crystal [9]. We also proposed adding molten quartz medium above the BGO crystal in the transverse modulation sensing head [10] to improve the electric field uniformity in the crystal. In the above study of our research group, quartz glass has dual functions as both an electric field equalizer and a voltage divider.

Although to employ a dielectric voltage divider is an effective method to increase OVS HWV, it still has some shortcomings. First of all, $FS_6$, as insulation and voltage-divided medium, is not environmentally friendly and has not adapted to the development requirements of green economy [11]. The stability of gas medium is insufficient, and the surface charge discharge [12], corona [13], plasma [14] and so on generated by gas medium seriously affect the stability of measurement. It is difficult to improve the space electric field uniformity by pressure equalizing ring or other means. Using solid partial pressure medium such as quartz glass or ceramic also has problems. To ensure the light path stable some curing glue is usually used for hard connection between the partial pressure medium and the BGO crystal. Due to the thermal expansion coefficient, the temperature change in the BGO and the quartz glass generates additional heat stress birefringence [15-19] degrading measurement accuracy. A linear OVS [20-22] based on an image demodulation can alleviate this effect due to the large measurement range, the high signal-to-noise ratio, and the measurement results are independent of optical power. However, because linear birefringence directly affects the electro-optic phase delay signal, the OVS based on image demodulation cannot completely eliminate the effect of linear birefringence.

In this paper, a new method is proposed to improve the HWV of the BGO crystal by adjusting the direction of electric field along the optical path. The high voltage electrode and the ground electrode adopt the center symmetrical heterogeneous electrode pair instead of the mirror symmetrical parallel plate electrode structure. The direction of the electric field on the light path is not perpendicular or parallel to the light propagation direction but has a certain angle. The electric field generated by the electrode is emulated by the finite element method and the electro-optic modulation is calculated by the coupling wave theory. The results show that the sensitivity of the light intensity signal to the measured voltage can be changed by adjusting the direction of the electric field, so as to increase the HWV of BGO crystal. The experimental results show that the dependence between the HWV and the direction of the average electric field generated by the center-symmetric electrode is consistent with the trend of theoretical prediction. Due to the angle errors of the direction of the incident light, the direction of polarization and the direction of crystal cutting, the actual adjustable maximum half wave voltage is less than the theoretical predicted value. However, further enhancement of the HWV is expected with more sophisticated optical-mechanical components. To ensure a 0.2 accuracy class, the angle error of incident light, polarization direction and electric field direction is required to be less than 3.9'.

## II. Theoretical analysis of electro-optic effect in arbitrary electric field direction

The refractive index ellipsoid theory is usually used to describe the electro-optic effect of transverse or longitudinal modulated BGO crystals. The schematic diagram of transverse modulation is shown in Figure 1. For an arbitrary direction of electric field, that is, the angle between the direction of incident light and the direction of electric field is not 0 degree or 90°, which is not can be described by the refractive index ellipsoid theory. In this paper, the electro-optic effect coupling wave theory [23-25] is used to calculate the electro-optic effect generated by the center-symmetric electrode.

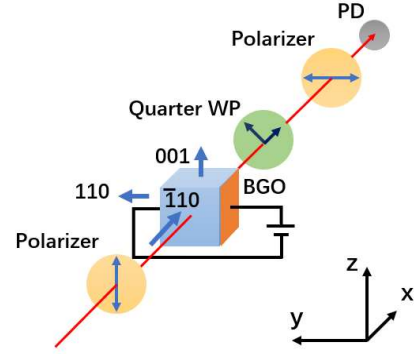

Fig. 1 110 Schematic diagram of transverse modulation of the BGO crystal, the direction of light is $\bar{1}10$, and the direction of electric field is 110

As shown in Figure 1, the spatial coordinate system used in the finite element method is different from the crystal coordinate system. The x, y and z directions of the former correspond to the crystal oration $\bar{1}10$, 110 and 001 respectively. When calculating with the coupling wave theory, the spatial coordinate system needs to be converted to crystal coordinate system. Let the axis coordinates of the crystal be X, Y and Z. The transformation relationship is shown in Equation (1):

$$X = \frac{1}{\sqrt{2}}(y - x)$$
$$Y = \frac{1}{\sqrt{2}}(y + x) \quad (1)$$
$$Z = z$$

Considering the truth that the direction and magnitude of electric field are not uniform on the light propagation path, this paper adopts the calculation method in literature [26]. A section with a sufficiently small length along the path is approximately regard as an electric field uniform region, and described by Jones matrix. The input and the output polarized light of this section satisfy Equation (2) and (3).

$$\begin{bmatrix} E_1^{j+1} \\ E_2^{j+1} \end{bmatrix} = A_j \begin{bmatrix} E_1^j \\ E_2^j \end{bmatrix} \quad (2)$$

$$A_j = \begin{bmatrix} \cos(\mu^j r) + i\gamma^j \sin(\mu^j r)/\mu^j & -id_1^j \sin(\mu^j r)/\mu^j \\ -id_3^j \sin(\mu^j r)/\mu^j & \cos(\mu^j r) + i\gamma^j \sin(\mu^j r)/\mu^j \end{bmatrix} \quad (3)$$

Where $\mu^j$, $\gamma^j$, $d_1^j$, $d_3^j$ are the effective electro-optical phase factors, $r$ is the light length of the tiny path, and $I$ is an imaginary symbol. The $j$ is Jones matrix of electro-optic effect generated by this path. The electro-optical effect on the entire optical path can be expressed as:

$$\begin{bmatrix} E_1^n \\ E_2^n \end{bmatrix} = Q_{\lambda/4} \prod A_j \begin{bmatrix} E_1^0 \\ E_2^0 \end{bmatrix} \quad (4)$$

$$Q_{\lambda/4} = \frac{1}{\sqrt{2}} \begin{bmatrix} 1 & i \\ i & 1 \end{bmatrix} \quad (5)$$

Where $Q_{\lambda/4}$ is the Jones matrix of 1/4 wave plate, which is used to move the static working point to the approximate linear region, so that the output optical power meets the approximate linear relationship with the electro-optic phase delay in a small angle range.

The experimental BGO crystal size is 10mm*10mm*10mm, with transverse modulation. The temperature is 27 degrees Celsius, and the central wavelength of the semiconductor laser source is 976nm. An AC voltage with a peak value of 500V is applied to the electrode. According to the ellipsoid formula of refractive index, phase delay, output light intensity and HWV are respectively:

$$\delta = \frac{2\pi}{\lambda} n_0^3 r_{41} V \frac{h}{d} \quad (6)$$

$$I_{out} = \sin^2[(\delta + \frac{\pi}{2})/2] \quad (7)$$

$$V_{\lambda/2} = \lambda d / 2 n_0^3 r_{41} h \quad (8)$$

The calculated HWV is 47.06kV, and the peak voltage is 500V in the experiment, which is 1.56% of the HWV. Therefore, the working point is in an approximate linear region.

Assuming that the electric field is 5e4V/m and the direction changes from y direction (110) to x direction ($\bar{1}$10), the angle between the electric field direction and y direction is assumed to be $\theta$. The relationship between $\theta$ and the HWV is shown in Fig. 2. When the electric field direction is close to the light direction, the HWV will increase infinitely, that is, the electro-optic phase delay will approach zero. Table I shows the change of HWV as the direction of electric field approaches 90 degrees. It should be pointed out that this mode is different from the standard longitudinal modulation mode of the BGO, with the polarization or detecting direction along 100 or 010.

## II. The electric field regulation method in the crystal based on the centrosymmetric electrode

An optical voltage sensing head based on the BGO crystal usually adopts mirror symmetry parallel plate structure, and the electric field direction is parallel or perpendicular to the light propagation direction. The Fig.3(a) shows a central symmetry electrode structure with a length of 5mm in the x direction and 2mm in the y direction. With this electrode, the average direction of the electric field along the optical path in the crystal is 18.21 degrees. Due to an edge effect, the direction and the magnitude of electric field at the edge are largely different from that at the center, as shown in the Fig.3(b) and the Fig.4. The simulation parameters used in the electrostatic finite element simulation are as follows: the dielectric constant is 16.4, the electrode thickness is 0.1mm, the material is copper, the upper right electrode is a high voltage electrode, and the voltage is 500V.

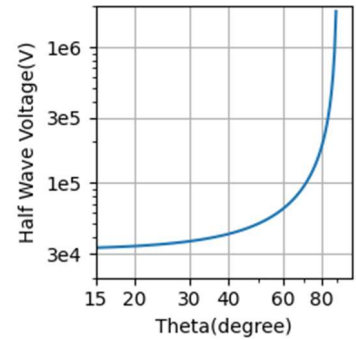

Fig. 2 The relation between HWV and electric field direction Angle obtained by theoretical calculation

Table I Theoretical values of HWV with same electrical field magnitude

| Theta(degree) | $\Delta I/\Delta V$(1/V) | HWV(kV) |
|---|---|---|
| 85 | 2.93e-6 | 534.75 |
| 86 | 2.35e-6 | 668.14 |
| 87 | 1.76e-6 | 890.54 |
| 88 | 1.17e-6 | 1.33e3 |
| 89 | 5.88e-7 | 2.67e3 |
| 89.9 | 5.88e-8 | 2.67e4 |
| 89.99 | 5.88e-9 | 2.67e5 |

The Fig.5 shows all the electrode structures studied in this paper in which the electrodes are divided into two types. When the electric field direction is close to the transverse modulation electric field, a copper foil electrode is used. When approaching the longitudinal modulation, the ITO transparent electrode is used to ensure that the polarized ray incident from the center of the crystal plane -110 without being blocked. The Table I shows the average electric field angle and the variation range of electric field angle with all electrode structures. It can be seen that the central symmetry electrode is an effective means to regulate the direction of electric field in crystal. Table II shows the HWV calculated based on the coupling wave theory with uniform electrical field along optical path. It can be seen that when the results in Table II is consistent to Table I, where the electrical field is assumed to be uniform in the whole optical path. However, the maximum HWV in Table III, namely 2.513e5 kV, is not a stable value in real world. Because the actual crystal cutting direction and the light propagation direction would have certain angle error within 1 or 2 degrees. It can be deduced from the Table I, if angle error is about 1degree, the HWV would decrease by 2 orders of magnitude. The following experimental results are in good agreement with these results.

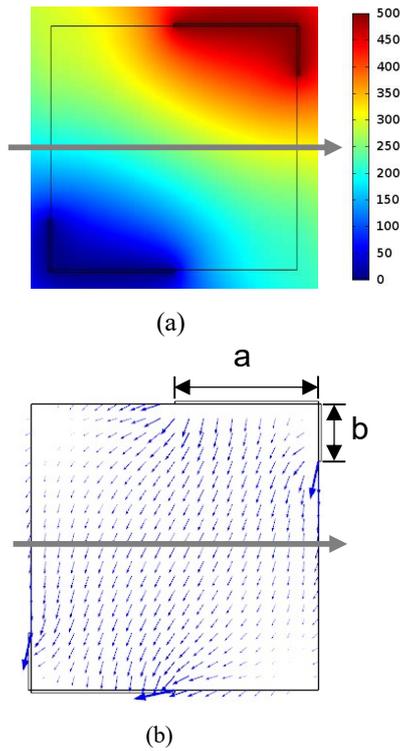

(a)

(b)

Fig 3 Finite element simulation analysis of electric field in BGO crystal based on asymmetric electrode (a) electric potential distribution and (b) electric field direction distribution

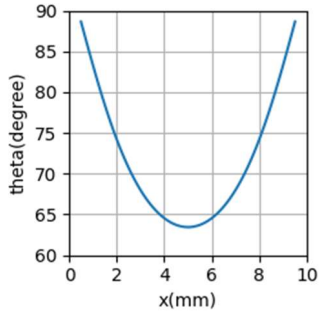

Fig. 4 Distribution of electric field direction on the light path, where $\theta$ is the Angle between electric field direction and y direction.

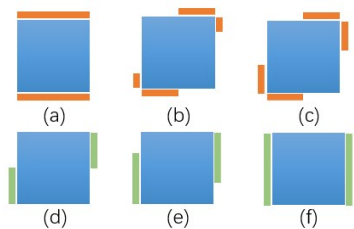

Fig. 5 Schematic diagram of transverse and longitudinal modulated electrode and center-symmetric electrode used in this paper. The ratio of electrode length a in x direction to electrode length b in y direction is (a)10:0 (b) 5:2 (c) 5:4 (e) 1:4 (f) 0:5 (g) 0:7 (f) 0:10, and the average electric field direction is: (A) 0degree, (b)18degree, (C)60degree, (e) 30degree, (f) 7.6degree, (g) 0.01degree

Table II Calculated electrical field distribution along the optical path

| Electrode length ratio a: b | Electric field direction: theta (degree) | | | |
|---|---|---|---|---|
| | Maximum | Minimum | Variation | Average |
| 10:0 | 0 | 0 | 0 | 0 |
| 5:2 | 1.35 | 26.56 | 25.21 | 18.21 |
| 5:4 | 2.33 | 37.80 | 35.47 | 30.13 |
| 0:5 | 67.67 | 48.21 | 19.46 | 60.39 |
| 0:7 | 89.65 | 80.71 | 8.9381 | 82.37 |
| 0:10 | 90.00 | 89.98 | 0.0188 | 89.99 |

Table III Theory values of HWV with emulated electrical field

| Electrode length ratio | Theta | $\Delta I$ | HWV(kV) |
|---|---|---|---|
| 10:0 | 0 | 0.01669 | 47.06 |
| 5:2 | 18.21 | 0.01236 | 63.57 |
| 5:4 | 30.13 | 0.01495 | 52.53 |
| 0:5 | 60.39 | 0.01047 | 75.02 |
| 0:7 | 82.37 | 0.00221 | 355.30 |
| 0:10 | 89.99 | 3.13e-6 | 2.513e5(**unstable**) |

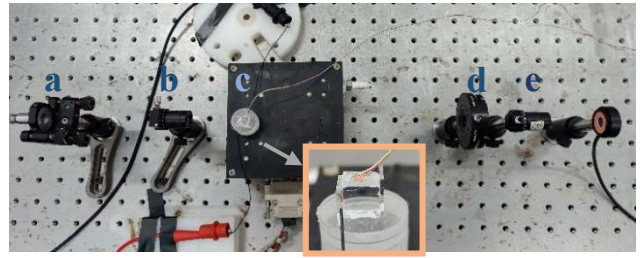

Figure 5 experimental device diagram (a) Tail fiber collimator of the semiconductor laser, installed on the optical adjustment frame (b) Glan Tayler prism, namely polarizer; (c) BGO cubic crystal, the passing surface is $\bar{1}10$, the electric field direction is 110. The electrode is ITO transparent electrode, and the wire is fixed on the ITO conductive surface with copper foil tape. Quartz glass sheet and semiconductor temperature control platform are below the crystal. (d)1/4 wave plate (e) Glan Taylor prism. (f) Silicon photodiode optical power meter, wavelength range 400-1100nm.

### III. Experimental results

The experimental set up is shown in Fig.5. The light source is a semiconductor laser diode produced by the Thorlabs, with a central wavelength of 976nm. The polarizer adopts the Gran Taylor prism with high extinction ratio. The analog signal generated by the optical power meter is recorded by the 16-bit acquisition card with 100MHz sampling rate of the NI Company, with model NI USB-6361. The 50Hz voltage signal is produced by a high voltage amplifier produced by the Pinzhi Company, and the model is HA-2400. Its signal input source is the signal generator produced by the Rigol with model DG1022. The BGO crystal is a cube with a side length of 10mm, and the cutting direction is 110, $\bar{1}10$ and 001. BGO crystals were grown by pull-up method at the Shanghai Institute of Ceramics, Chinese Academy of Sciences. A type K thermocouple is pasted directly above the crystal through copper foil tape and connected to the acquisition card for monitoring the temperature of BGO crystal. The copper foil electrode or ITO transparent electrode is fixed on the BGO crystal with UV curing adhesive.

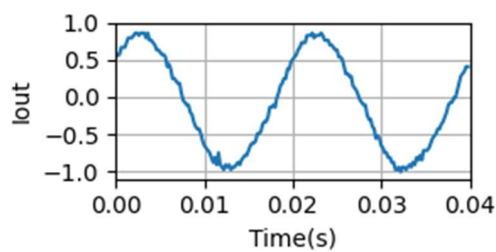

Fig. 6 Output optical power signal under standard transverse modulation. A sinusoidal signal with a modulated voltage of 50Hz and a peak voltage of 300V, generated by a high voltage amplifier.

Table IV Experimental measurement results of HWV

| Electrode length ratio a:b | Theta | $\Delta I/\Delta V$(uW/V) | HWV(kV) |
|---|---|---|---|
| 10:0 | 0 | 7.82e-3 | 49.32 |
| 5:2 | 18.21 | 4.90e-3 | 78.68 |
| 5:4 | 30.13 | 5.30e-3 | 72.74 |
| 0:5 | 60.39 | 4.85e-3 | 79.49 |
| 0:7 | 82.37 | 2.14e-3 | 180.16 |
| 0:10 | 89.99 | 8.94e-4 | 1.72e3 |

## IV. Results and Discussions

The Fig.6 shows the optical power signal in the standard transverse modulation mode, where the DC component is minus. The Table III shows the HWV values corresponding to different electrode structures, which is consistent with the trend of theoretical calculation results in the Table II. This shows that the coupling wave theory can describe the electro-optic effect better in arbitrary electric field direction.

The difference between the actual measured maximum regulated HWV and the theoretical calculated value with 89°, which is a stable HWV value as shown in the Table I, is 35.6%. For real experimental conditions, the unstable half wave voltage is difficult to realize because the following reasons:(1) the incident polarization direction is not strictly parallel to the crystal 110 direction, and (2) there is a small angle between the incident light direction and the normal incident direction. (3) there is an angle error in the crystal cutting direction. In the future, we plan to use precision angle rotating table and other equipment to reduce the angle error of crystal cutting direction and improve the maximum adjustable voltage.

According to the Fig.5, to make the HWV of the BGO crystal reach the line voltage of 110kV, namely the phase voltage of 63.5kV, to ensure the measurement accuracy of class 0.2, the HWV error should not exceed 0.127kV, and the Angle error should be less than plus or minus 0.065 degrees, namely 3.9 minutes. It can be achieved by existing optical and mechanical techniques, but new requirements for seismic performance of OVT are put forward.

## V. Conclusions

In this paper, a HWV regulating method for the BGO crystal based on an asymmetric electrode and a special directional modulation electric field is proposed. Through the calculating with electro-optic coupling wave theory, the HWV can be increased to 2.67e3 kV as a stable value by using center-symmetric electrodes. Experimental results are consistent with theory calculation, and the difference between them is 35.6%. The HWV can be further increased if the polarization direction of incident light is completely parallel to the crystal 110 direction and the light propagation direction is completely parallel to $\bar{1}10$ direction. Our proposed HWV regulation method only requires asymmetric electrodes for changing the direction of electric field in the crystal, without using additional voltage divided media. A theoretical analysis indicates that, If the HWV is adjusted to 63.5kV (line voltage 110kV), the accuracy of electric field Angle should reach 3.9'. The angle precision can be achieved by the existing optical technology and shock absorption technology. In order to achieve accurate regulation of HWV and ensure the stability of electro-optical phase signal, the uniformity of electric field in crystal needs to be further optimized.

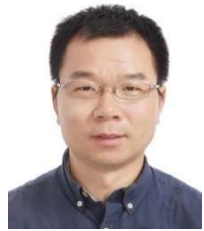

**Junfeng Chen** received the B.S. degree in Materials Science and Engineering from Zhengzhou University, P.R. China, in 2003 and the M.S. degree in Material Sciences, Ph.D. degree in Material Physics and Chemistry from Shanghai Institute of Ceramics (SIC), Chinese Academy of Sciences (CAS), in 2006, and 2015, respectively. He visited the Division of Physics, Mathematics and Astronomy (PMA) of California Institute of Technology as a Visiting Associate in 2016 sponsored by CAS.

He has been with Shanghai Institute of Ceramics, Chinese Academy of Sciences since 2006 and is currently working as an Associate Professor since 2016. His research interest includes the discovery and development of novel materials with scintillation and electro-optic effects, implementation of emerging high performance single crystals and functional devices for applications, such as voltage and electric field sensors, medical imaging instruments, homeland security inspection systems, well-logging and geological exploration, nuclear and particle physics experiments, etc.

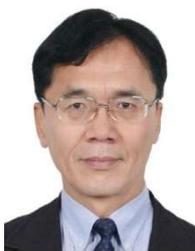

**Qifeng Xu** was born in Jinan, China, in 1959. He received the B.S. degree in electric power system and automation from the Shandong Institute of Technology, China, in 1983, and the M.S. and Ph.D. degrees in electric power engineering from North China Electric Power University, China, in 1986 and 1992, respectively. From 1993 to 1995, he was a Research Fellow with the Queen's University of Belfast, U.K.

From 1996 to 2007, he was a Senior Engineer with Motorola, U.K. Since 2008, he has been a Professor and a Ph.D. Tutor with the College of Electrical Engineering and Automation, Fuzhou University, Fuzhou, China. Since 2017, he has been a Visiting Professor with Longyan University, Fujian, China. His research interests include electric power system measurement and control technologies.

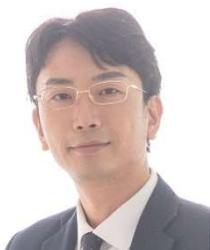

**Nan Xie** was born in Huainan, China, in 1985. He received the B.S. degree in Material Physics, the M.S. degrees in Material Physics and Chemistry and the Ph.D. degree in Optics from Nankai University, Tianjin, China, in 2007, 2010, and 2013, respectively. He was also co-cultivated at the Institute of Physics, Chinese Academy of Sciences from 2009 to 2013.

He has been with College of Electrical Engineering and Automation, Fuzhou University since 2013 and is currently working as an Associate Professor since 2021. His research interest includes optical voltage/current sensor, dielectric physics of linear electro-optic crystals, polarization optical detection technology and fiber optics technology, etc.

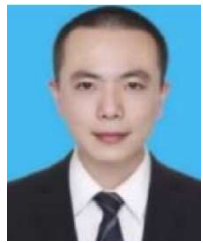

**Yifan Huang** was born in Fujian, China, in 1993. He received the B.S. and Ph.D. degrees in electric power systems from Fuzhou University, China, in 2015 and 2020, respectively.

Since 2020, he has been a Lecturer with the College of Electrical Engineering and Automation, Fuzhou University. His research interest includes optical transducer technology applied in the electric power systems.

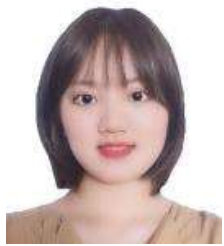

**Yifan Lin** was born in Fujian, China, in 1997. She received a bachelor's degree in electrical engineering from Guilin University of Electronic Technology, China, in 2020.

She is currently pursuing a master's degree in energy and power at Fuzhou University, Fuzhou, China. Her research interest includes optical voltage sensor technology and polarization optical detection technology.